\documentclass[%
 aip,
 apl,%
 amsmath,amssymb,
]{revtex4-1}

\usepackage[utf8x]{inputenc}

\usepackage[T1]{fontenc}

\usepackage{amsfonts}

\usepackage{array}

\usepackage{multirow}
\usepackage{eurosym}

\usepackage{pslatex}
\usepackage{graphicx}
\usepackage{color}
\graphicspath{{figures/}}
\usepackage{dcolumn}

\begin{document}

\title{Unprecedented High Long Term Frequency Stability with a Macroscopic Resonator Oscillator}

\author{S. Grop}
\affiliation{FEMTO--ST Institute, Time and Frequency Dpt., UMR 6174 CNRS-UFC-ENSMM, 32 av.~de l'Observatoire, 25044 Besan\c{c}on Cedex, France}
\author{W. Sch\"afer}
\affiliation{TimeTech GmbH, Curiestrasse 2, D-70563 Stuttgart, Germany}
\author{P.Y. Bourgeois}
\author{N. Bazin}
\author{Y. Kersal\'e}
\affiliation{FEMTO--ST Institute, Time and Frequency Dpt., UMR 6174 CNRS-UFC-ENSMM, 32 av.~de l'Observatoire, 25044 Besan\c{c}on Cedex, France}
\author{M. Oxborrow}
\affiliation{National Physical Laboratory, Queens Road, Teddington, Middlesex, TW11 0LW, UK}
\author{E. Rubiola}
\author{V. Giordano}
\affiliation{FEMTO--ST Institute, Time and Frequency Dpt., UMR 6174 CNRS-UFC-ENSMM, 32 av.~de l'Observatoire, 25044 Besan\c{c}on Cedex, France}
\email[giordano@femto-st.fr]{giordano@femto-st.fr}

\date{\today}

\begin{abstract}
This article reports on the long-term frequency stabilty characterisation of a new type of cryogenic sapphire oscillator using an autonomous pulse-tube cryocooler as its cold source. This new design enables a relative frequency stability of better than $4.5\times 10^{-15}$ over one day of integration. This represents to our knowledge the best long-term frequency stability ever obtained with a signal source based on a macroscopic resonator.

\end{abstract}

\maketitle

%\clearpage

An oscillator consists of a resonator in closed loop with a sustaining amplifier that compensates for losses. 
The frequency stability is limited by the noise of the amplifier through the Leeson effect \cite{rubiola-phase-noise} and by the fluctuation of the resonator's natural frequency. It turns out that the stability of the resonator is by far the most important parameter that determines the long-term stability, while the noise of the sustaining amplifier affects only the phase-noise and the short-term stability. When the very-long-term stability is the most important parameter, as in timekeeping and in radionavigation systems, atomic resonances 
are the only viable frequency references. In this case, a flywheel oscillator is frequency locked to the atomic resonance. On the other hand, macroscopic-cavity resonators show several advantages versus the atomic resonators because of their simplicity, reliability and power-handling capability. Higher power results in higher signal-to-noise ratio, and ultimately in low phase noise and high short-term stability.
 Ultimate stability in the range of $1-10^{6}$ s measurement time is of paramount importance in physical experiments involving long averaging, and of course in radioastronomy.\\
 
 In this paper we demonstrate for the first time a microwave oscillator based on a macroscopic resonator with a frequency stability at long integration times that is competitive with those of classical microwave atomic clocks. \\

Microwave Cryogenic Sapphire Oscillators (CSO) exhibits the higest short-term stability, attaining parts in $10^{-16}$ near 10 s integration time \cite{chang00,hartnett06-apl,watabee06-eftf}. A CSO incorporates a cryogenic whispering gallery mode resonator made in sapphire which provides a Q-factor as high as $1\times 10^{9}$ at $4.2$~K.
In all functional realization up to now, the resonator is immersed in a liquid-helium bath and maintained at its optimum temperature (generally around 6~K), where its thermal sensitivity nulls to first order. CSOs have been used as local oscillators for atomic fountains clocks\cite{santarelli99-prl}  and for fundamental physical experiments as Local Lorentz Invariance tests \cite{wolf03}. It is also planned to implement such oscillators in Deep Space Network ground stations to improve the tracking of space vehicules and in VLBI observatory for better data correlation. For these last applications, the use of liquid helium is inconvenient and a change of technology is needed.
 We recently validated in the frame of a European Space Agency research contract, an new instrument: ELISA based on a CSO operating in a specially designed cryocooler. The detailed design and preliminary characterisations can be found in \cite{rsi10-elisa,ell10-elisa}. This CSO is associated with a frequency synthesis delivering round frequencies, i.e. 10 GHz, 100 MHz and 5 MHz. The ELISA's relative frequency stablity is better than $3\times 10^{-15}$ for $1\mathrm{s}\leq \tau \leq 1,000$s and  can operate continuously for two years without maintenance. As an additional benefit, this new type of CSO presents an unprecedented frequency stability at long integration times: 
 $4\times 10^{-15}$ over one day without any clearly observed drift.\\

\begin{figure}[ht!]
	\centering
	\includegraphics[scale=0.7]{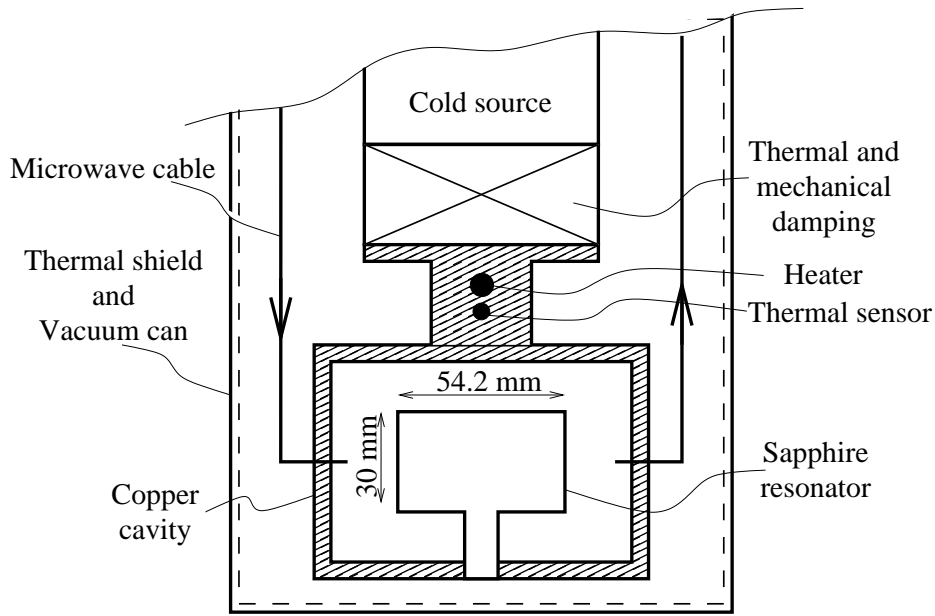}
	\caption{\it{Scheme of the synthesis used to generate the 10GHz, 100 MHz and 5 MHz signals}}
	\label{fig1}
	\end{figure}

The resonator consists of a Crystal System HEMEX grade single-crystal sapphire, 54.2 mm diameter and  30 mm thickness with a 10 mm diameter spindle allowing a stable mechnical clamping in the center of a gold plated copper cavity. This assembly as schematised in figure  \ref{fig1}
is fixed on the experimental cold plate of a pulse-tube cryocooler. A special \it{soft}\rm\ thermal link  and a thermal ballast were designed in order to filter the vibrations  and the temperature modulation at $1.4$ Hz induced by the gas flow in the cryocooler.

Using a mechanical model, the g-sensitivity of the frequency of the resonator's operating mode was determined to be $3.2\times 10^{-10}/$g \cite{eftf08-g-sensitivity}. Phase noise measurements about the pulse-tube's cooler frequency of reciprocation (~1 Hz), thereupon demonstrated that the residual displacement of the resonator is less than $2~\mu$m at the cryocooler cycle frequency \cite{ell10-elisa}.  
 The resonator temperature is stabilized  at 6.1~K with overal temperature rms fluctuations of $\pm 1$~mK. At 6.1~K, the first order resonator frequency thremal sensitivity is nulled. The residual second-order thermal sensitivity is $1.6\times 10^{-9}$~K$^{-2}$.  The CSO is completed by an external sustaining circuit and two servos to stabilize the power injected into the resonator and the phase lag along the sustaining loop. 

\begin{figure}[ht!]
	\centering
	\includegraphics[scale=0.5]{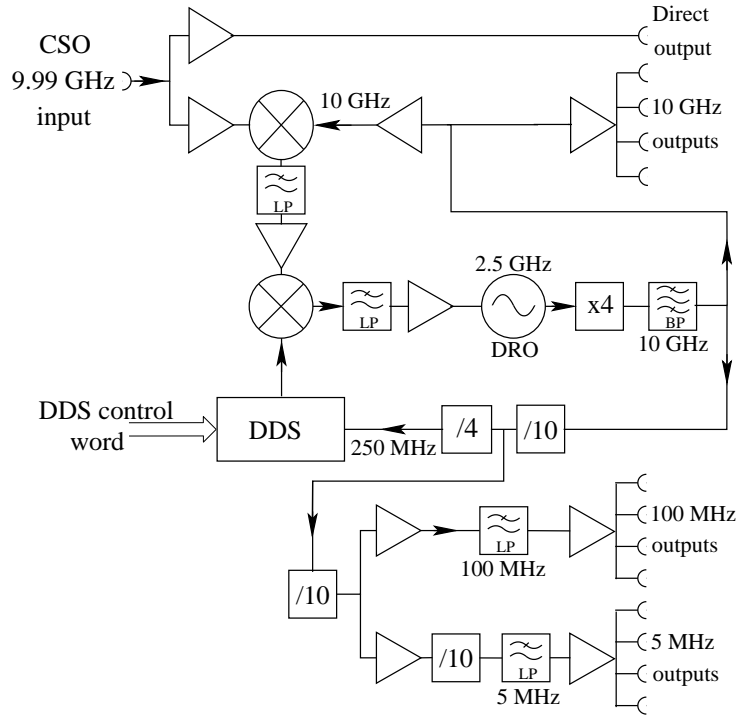}
	\caption{\it{Scheme of the synthesis used to generate the 10GHz, 100 MHz and 5 MHz signals}}
	\label{fig2}
	\end{figure}

The mechanical tolerances in the resonator machining induce an uncertainty in the actual resonator frequency  of $\pm~3.5$~MHz. The resonator was designed to operate on the  $WGH_{15,0,0}$ whispering gallery mode at $9.99$~GHz. The intentional $10$ MHz frequency offset from the 10~GHz round frequency was chosen to permit to compensate for the resonator frequency uncertainty by using a low noise Direct Digital Synthesizer (DDS). The actual resonator frequency measured at 6.1~K is 9.989,121~GHz. The  anatomy of the frequency synthesis chain is shown in the  figure  \ref{fig2}. \\

A 2.5 GHz Dielectric Resonator Oscillator (DRO) choosen for its low phase noise is frequency multiplied by 4 and mixed with the CSO's signal. The resultant 11 MHz beatnote is compared to the signal coming from a DDS. The resulting error signal is used to phase lock the DRO to the CSO.
Frequency dividers complete the system to generate the 100 MHz and 5 MHz frequencies from the 10 GHz signal.

To evaluate the Elisa frequency stability on a large integration times range, two other frequency references have been used: i) we implemented a second CSO: Aliz\'ee but cooled in a liquid-helium dewar and equipped with the same frequency synthesis. The Elisa and Aliz\'ee resonators are almost identical and were machined from the same sapphire boule. Aliz\'ee's resonator is placed in a vacuum can immersed in a 100 litre liquid-helium dewar. 
 ii) An hydrogen maser which does not include an automated cavity tuning. The latter instrument was placed in the same room as the two CSOs. This room is not temperature controlled.
The results of the two comparisons are summarized in the figure  \ref{fig3}:

\begin{figure}[h]
\centering
\includegraphics[scale=1.1]{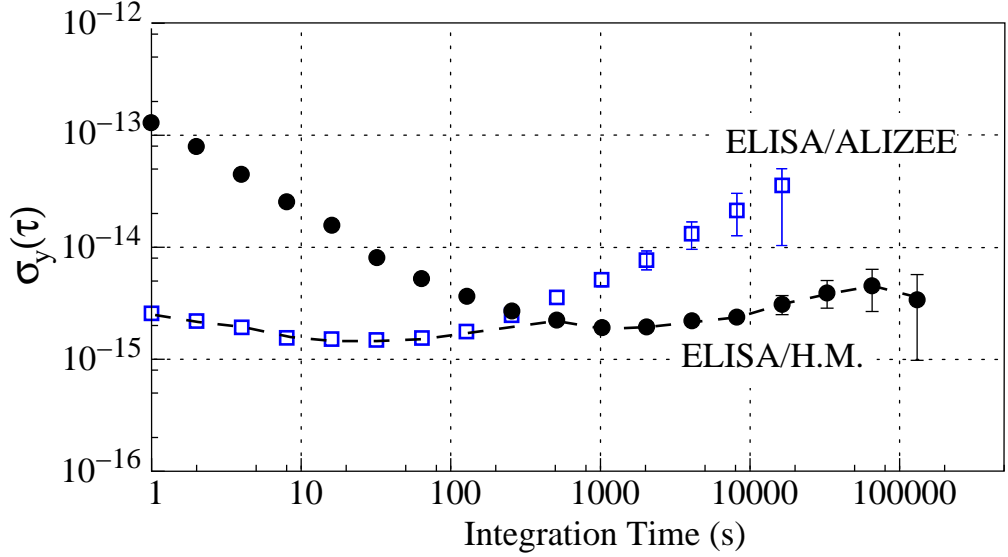}
 \caption{ \it{Relative frequency stability as measured by beating ELISA with another CSO (\textcolor{blue}{$\square$}) and the hydrogen maser (\Large$\bullet$\normalsize). The broken line represents a conservative ELISA's frequency stability estimation from 1s to more than 1 day. }}
 \label{fig3} 
\end{figure}

\begin{figure}[ht!]
	\centering
	\includegraphics[scale=1]{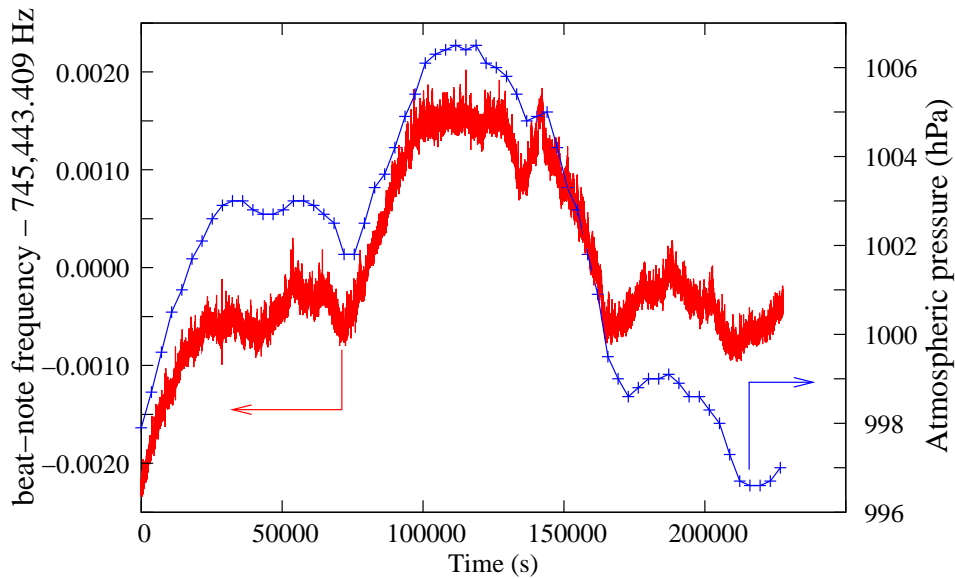}
	\caption{\it{Beatnote frequency (red curve) and atmospheric pressure (blue curve) as a function of time}}
	\label{fig4}
	\end{figure}

The short term frequency stability of the synthesized signals has been evaluated at 10 GHz by beating the Elisa's and Aliz\'ee's 10 GHz outputs. The Aliz\'ee output was intentionally frequency shifted by acting on the DDS command in order to get a 200 kHz beatnote. This beatnote was directly counted with a gate time $\tau=1$ s. Then the relative frequency deviation was calculated for the different intregration times by grouping the $1$ s data. We used an Agilent 53132A counter. This instrument has a specific stastical procedure giving a result $\sigma_{\Lambda}(\tau)$ which slightly differs from the true Allan deviation $\sigma_{y}(\tau)$\cite{rubiola-rsi05,dawkins07}. Nevertheless for the integration times we consider here, it gives an overestimated relative frequency deviation with respect to $\sigma_{y}(\tau)$. Moreover no data post-processing has been done and we did not divide the result by $\sqrt{2}$: a pratice which is generally adopted when comparating two almost equivalent oscillators. The results presented here are thus conservative.\\
The short-term frequency stability is limited by a white frequency noise process $\sigma_{\Lambda}(\tau)\approx 3\times 10^{-15}~\tau^{-1/2}$, which we attribute to the noise of the Pound servo used to stabilize the phase along the sustaining loop. At long term a random walk or a frequency drift alters the frequency stability. The implementation of Elisa was finalized in October 2009 and since it has been continuously running apart from two short periods of time, firstly for implementing an optimised rotary valve for the cryocooler and secondly after a general electrical breakdown arrising in our laboratory. After each stop, Elisa was simply switched on and recovered its optimal temperature in about 8 hours. Conversely, Aliz\'ee was stopped a number of times because of limited supplies in liquid helium. Moreover, for a long measurement compaign the dewar needs to be refilled  every 10 days. During this 6-months period, a number of measurements were realised, demonstrating that the long-term behavior of the beatnote depends substantially on the status of Alizee and on the environmental perturbations. As an example, figure  \ref{fig4} 
shows the correlation between the beatnote frequency and the local atmospheric pressure. The liquid helium temperature of evaporation depends on the atmospheric pressure with a sensitivity equals to 1mK/mbar near 4.2 K. Temperature exchange by radiation between the vacuum can and the Aliz\'ee resonator takes place and perturbs the mode frequency leading to long term instability.\\

The long-term stability has been evaluated at 100 MHz by using the hydrogen maser as reference. The relative frequency stability has been computed from the phase difference data averaged over a sampling periode of 1s. The data were taken continuously during more than 5 days and the Allan deviation was computed.  The result is shown in figure \ref{fig3} (black bullets). The maser short-term instability limits the measurement for $\tau \leq 500$ s, but for longer integration times, it is obvious that Elisa is far better than Aliz\'ee.  
The maximum frequency instability, i.e. $4.5\times 10^{-15}$ arises near 1 day. It is likely that the hydrogen maser itself significantly contributes to this frequency instability. Indeed, its residual sensitivity has been measured to be $1.4\times 10^{-14}/$K which is far from negligible given that the daily variation of the temperature in the laboratory was typically a few degrees Celsius.  \\

\begin{figure}[h!!!]
\centering
\includegraphics[scale=1]{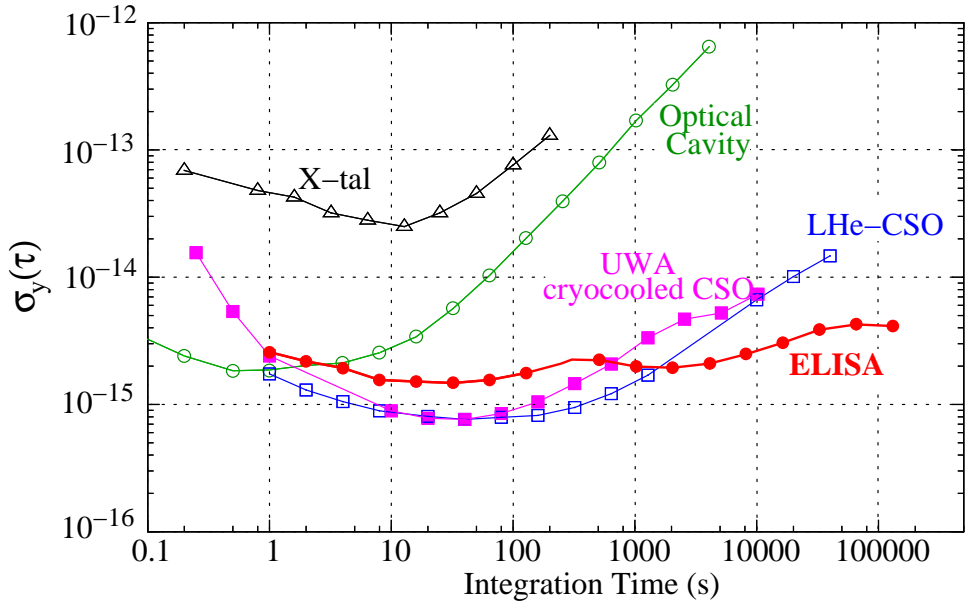}
 \caption{ \it{\textcolor{red}{\Large$\bullet$}: Estimated ELISA's relative frequency stability compared to some other frequency standards based on macroscopic resonator. 
$\triangle$: 5 MHz quartz oscillator \cite{salzenstein2010-ell}; 
\textcolor{green}{\Large$\circ$}: Laser stabilized on an ultra-stable optical cavity \cite{webster08};  \textcolor{magenta}{$\blacksquare$}: UWA cryocooled CSO \cite{nand10}; 
\textcolor{blue}{$\square$}: UWA liquid-helium cooled CSO \cite{hartnett06-apl}. 
NB: To be consistent with our own procedure to evaluate the frequency stability, the published results for which the two-equivalent-oscillators hypothesis has been assumed have been multiplied by a factor $\sqrt{2}$. }}
\label{fig5} 
\end{figure}

Elisa presents the highest frequency stability over one day ever obtained with an oscillator based on a macroscopic resonator. To illustrate this point the figure \ref{fig5} compares the previous Elisa relative frequency stability to the best ever published ultra-stable oscillator performances.\\

%\clearpage
\bf{Acknowledgements}\rm~: 
This work was supported by the European Space Agency (ESA).

\end{document}